\documentclass[sigplan,10pt]{acmart}
\usepackage{siunitx}
\usepackage{subcaption}
\usepackage{tikz}
\usetikzlibrary{shapes.geometric, arrows}
\usepackage{hyperref}
\usepackage{cleveref}
\usepackage{bm}
\crefformat{section}{\S#2#1#3} 
\crefformat{subsection}{\S#2#1#3}
\crefformat{paragraph}{\S#2#1#3}
\crefformat{subsubsection}{\S#2#1#3}
\usepackage{enumitem}
\usepackage{graphicx}
\renewcommand\footnotetextcopyrightpermission[1]{} 

\AtBeginDocument{%
  \providecommand\BibTeX{{%
    \normalfont B\kern-0.5em{\scshape i\kern-0.25em b}\kern-0.8em\TeX}}}
    
\begin{document}

\title{Slowing Down for Performance and Energy: An OS-Centric Study in Network Driven Workloads}

\author{Han Dong, Sanjay Arora*, Yara Awad, Tommy Unger, Orran Krieger, Jonathan Appavoo\\
  \textit{Red Hat*, Boston University}
  }

\settopmatter{printfolios=false}

\begin{abstract}

This paper studies three fundamental aspects of an OS that impact the performance and energy efficiency of network processing: 1) batching, 2) processor energy settings, and 3) the logic and instructions of the OS networking paths. A network device's interrupt delay feature is used to induce batching and processor frequency is manipulated to control the speed of instruction execution. A baremetal library OS is used to explore OS path specialization. This study shows how careful use of batching and interrupt delay results in 2X energy and performance improvements across different workloads. Surprisingly, we find polling can be made energy efficient and can result in gains up to 11X over baseline Linux. We developed a methodology and a set of tools to collect system data in order to understand how energy is impacted at a fine-grained granularity. This paper identifies a number of other novel findings that have implications in OS design for networked applications and suggests a path forward to consider energy as a focal point of systems research.

\end{abstract}

\maketitle
\pagestyle{plain} 

\section{Introduction}
\label{sec:intro}

There has been a large body of work in systems research focused using OS path specialization techniques to accelerate network applications~\cite{ix, arrakis, zygos, shenango, rumpkernel, aliraza, unikernels, scalingmcdfacebook, arachne, mtcp, sandstorm, affinityaccept, flexnic, mica}. While performance is the main motivator behind these systems, we find their impact on energy is not as clearly understood. Our work seeks to start filling this gap by offering a detailed OS-centric study of performance-energy in network applications under different OS structures. 

There are three fundamental aspects of an OS that impact the performance and energy efficiency of network processing: 1) batching, 2) processor energy settings, and 3) the logic and instructions of the OS networking paths. Delaying packet processing improves overall software stack efficiency as system overheads such as interrupt processing, OS book-keeping, and cache misses are amortized or eliminated by the batched handling of packets. However, the benefits of batching are typically weighed against its impact on workload latency~\cite{mootaz}. Similarly, processor energy settings impacts the efficiency of network processing by trading off instruction execution speed with a reduction in energy use. In addition, batching and processor energy settings interacts with the software stack, its policies, and workload performance requirements to impact the energy saved by processor sleep states during idle periods between packet arrivals. Lastly, specializing OS paths offers the chance to handle packets with optimized OS logic and data structures, thus improving overall network processing efficiency. In this paper, our goal is to study how all three of these aspects interact together to impact network processing performance and energy use. 


A network interface controller's (NIC) interrupt delay feature~\cite{intelitr} is used to induce batching. A processor's Dynamic Voltage Frequency Scaling (DVFS)~\cite{cpufreq_governor} feature is used to control its frequency and energy setting to explores trade-offs in execution speed and energy use. An open sourced library OS (EbbRT~\cite{ebbrt}), ported to run baremetal, is used as a platform to contrast against a general purpose Linux to study how different OS structures are impacted by the two mechanisms listed.

While our data-driven study reveals a wealth of results in \cref{sec:exp} with different impacts on system design, below, we summarize three of the main example findings and suggest implications for how OSes can improve energy efficiency while supporting high performance network applications:
\begin{enumerate}[topsep=2pt, partopsep=0pt]
     \item \textbf{Finding:} By manually setting both DVFS and NIC interrupt delay values in an exhaustive search, we were able to find optimal performance and energy efficiency points in both OSes, e.g. rather than setting DVFS and interrupt delay values using its default policies, Linux configured with manually searched values can improve tail latency by 2X or energy savings by 55\% in a network driven TPC-C style transactional database workload (\cref{fig:mcdsilo_overview}). \textbf{Implication:} Careful coordination among different hardware features should be used towards common objectives to explore new trade-offs that achieve even further efficiencies over today's isolated policies.
     \item \textbf{Finding:} Polling not only improves latency, but can be made energy efficient (for small payloads) under specialized OS paths using slowed DVFS. For example, in \cref{sec:mcd:poll}, we show 27\% improved tail latency for a key value store with 35\% less energy than when interrupts are used and 11X improvement in energy efficiency for a closed loop benchmark in \cref{sec:closed_loop:poll}). For other workloads that involve larger payloads or more application work, polling can result in negligible performance improvement while consuming dramatically more energy. For example in \cref{fig:mcdsilo_overview}, a baremetal library OS using poll achieved best case tail latency but used over 60\% more energy than Linux. \textbf{Implication:} Specialized systems that use polling to achieving low-latency~\cite{sandstorm, ix, arrakis, zygos, shenango, ebbrt, seuss, arachne, mtcp, whenpollisbetter} can be made energy-efficient with careful use of DVFS, further, these results suggests the importance for energy aware OS-level optimizations that can switch between poll and interrupt-driven IO in response to changes in demand and workload behavior. 
     \item \textbf{Finding:} Exploiting OS path specialization via a baremetal library OS yields improvements in not only performance but also energy over Linux (88\% energy efficiency improvement for NodeJS webserver in \cref{fig:closed_loop_overview}, 2X throughput for memcached in \cref{fig:mcd_overview2}, and TPC-C in \cref{fig:mcdsilo_overview}). Surprisingly, even in application-heavy workloads, OS path specialization can still result in significant energy savings (up to 85\% in \cref{fig:mcdsilo_overview}). \textbf{Implication:} There is enormous value to adopting specialized OS or path specialization (even in general purpose OSes) beyond virtual environments for both performance and energy efficiency.
     
     
\end{enumerate}

In order to arrive at these findings, we conducted an extensive experimental study over the two OSes with thousands of experimental combinations resulting in a dataset over 5 TB. Given this large dataset, we developed a methodology and visualization tool to help us identify the performance-energy trade-offs in a fine-grained manner and to understand the causal relations between the hardware mechanisms and its impact on different OS structures. We believe the methodology, dataset, and tools are all contributions to OS research in systems and will help other researchers to develop new insights of the impact of OS changes on energy use. We plan to open source both the dataset and the tools. 

In order to better reason about data we've collected, we begin by constructing a simple generic timeline of how packets are processed in a typical system in \cref{sec:workflow}, we then discuss performance and energy in~\cref{sec:slowdown} in the context of how we break this timeline down. Next, we discuss our experimental and software setup in \cref{sec:exp_setup} and then present our experimental data in \cref{sec:exp}. Related work is discussed in ~\cref{sec:related} and  we conclude in \cref{sec:dis}.

\section{Processing Break Down}
\label{sec:workflow}

From an OS perspective we break down network driven processing into stages that allows us to organize and reflect the OS and application interaction with the workload request timeline. This break down is illustrated in figure~\ref{fig:timeline} and shows a generic set of stages that all requests must go through. \footnote{Although this is drawn and discussed from the perspective of a single core, our analysis and evaluation assumes that multiple cores could be concurrently used to shorten servicing times.}


\subsection{Quiescent}
\label{sec:workflow:Quiescent}
Given the packet and transactional nature of network driven services, a quiescent period, in which no requests are present at the server, precedes activity on the server. The nature of the workloads drive the length of quiescent period and the nature of the work itself required to service the request. Network services tends to fall into two broad categories, Open and Closed loop~\cite{openverclosed}.   

\subsubsection{Open Loop:}
\label{sec:workflow:openloop}
In an open loop scenario like a Memcached workload, the external request rate induces an inter-arrival gap that will drive the quiescent period -- longer at lighter loads (lower queries-per-second (QPS)) and shorter at heavier loads (higher QPS). The arrival rate can largely be considered independent of the time required to service a request. Providers often set a Service-level Agreement (SLA) target, such as some percentage of requests to be completed under a stringent time budget, and there has been a wealth of research in using these SLA headrooms to lower datacenter energy use mainly by decreasing processor frequencies~\cite{Dynamo, SmoothOperator, oldi-pegasus, adrenaline, heracles, energyproportion, warehouse-power}.

\subsubsection{Closed Loop:}
\label{sec:workflow:closed_loop}
Examples of closed loop workloads are snapshotting a database to a remote server, video streaming or a middle tier service within a data center~\cite{Barroso:2009:DCI:1643608, oldi-study, oldi-pegasus, warehouse-power, energyproportion, WebSearch}.  The work to be done is a sequence of requests that have an inter-dependency on each other. Specifically, the arrival of the next request depends on how fast it takes to service the current request. From a server's perspective, the quiescent period will be bounded by time to transmit both the request and the reply, as well as the time on the client to generate the next request. In the closed loop scenario, one would like the server to complete every request quickly so that the overall time to complete a task is minimized and ideally use less energy in the process. 
\begin{figure}
\vspace{-0.20in}
\centering
\includegraphics[width=1\columnwidth]{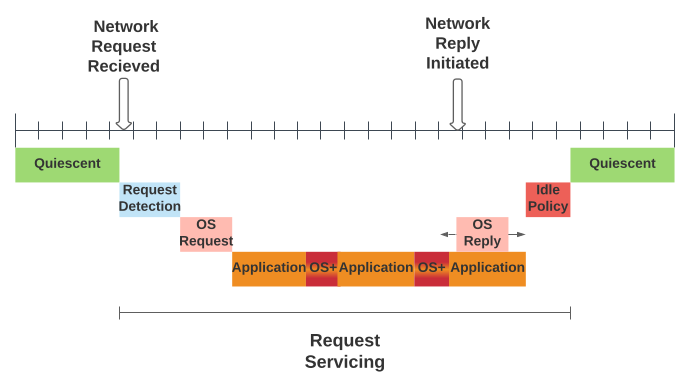}
\vspace*{-10mm}
\caption[]{Application processing request timeline. Quiescent are periods between packet arrivals. Request Servicing includes all software components.}
\label{fig:timeline}
\vspace{-0.10in}
\end{figure}
\subsection{OS Request Detection}
Fundamental to any operating system is how it detects and schedules processing in response to IO device activity.  At the two extremes are interrupt and poll driven detection.  

\subsubsection{Interrupt driven IO}
\label{sec:workflow:interruptio}
Using interrupts has three important implications: 1) it can be used to wake a processor from a halted state, which the OS entered to sleep the processor previously, in response to external activity, 2) allow an OS to arbitrate processing across competitive devices in a multi-programmed/multi-device setting, 3) interrupts have inherent performance costs associated with them --  latency in starting to handle a request, either because of the costs associated with preempting work~\cite{whenpollisbetter} or Intel c-state exit penalties\cite{cpuidle_policy}. This can also have a negative impact on the instruction efficiency, such as Instructions Per Cycle (IPC), due to induced micro-architecture hazards such as the inability to pre-fetch or speculatively execute across an interrupt.

\subsubsection{Hybrid driven IO}
\label{sec:workflow:hybridio}
A general purpose OS typically exploits some form of hybrid IO strategy alternating between interrupts and polling when servicing high speed NICs. A common strategy is to use interrupts when the load is low and switch to polling when load is high and back to interrupts when load reduces.  A general purpose OS, even under sustained high load, bounds the poll phase to avoid starving other devices and software. Linux's New API (NAPI)\cite{NAPI} framework implements this hybrid scheme.



\subsubsection{Poll driven IO}
\label{sec:workflow:pollio}
In contrast, specializing an OS to support the execution of a single application can explore more extreme strategies like aggressive polling. Most modern NICs devices expose a cache-friendly interface that permits the processors to read a per-core memory address to determine if the device has received data that requires processing by the core.  This allows software to directly poll the device and initiate software handling without an interrupt.  This approach reduces latency and other performance penalties associated with interrupt driven IO but requires a busy CPU. As a result, there have been a large body of work to address latency sensitive applications through the judicious use of polling~\cite{ix, arrakis, zygos, shenango, ebbrt, seuss, arachne, mtcp, whenpollisbetter, sandstorm}. In the extreme, a customized OS path supporting a single application can run a poll loop on every core to constantly check for work, conduct the work and then go back to polling for new work and thus never halting the processors.\footnote{It is worth noting processors also have the ability to halt in a way that an update to a cache line will awaken it, there exists the possibility of implementing the poll in combination with sleep states. We do not explore this possibility, leaving it for future work.} 







\subsection{OS Request Processing}
\label{sec:workflow:osreqproc}

Once the OS detection mechanism identifies the NIC has data to process, several components of OS functionality must be run in accordance with the execution model of the OS. Normally, a network stack parses the packet header and eventually passes the payload to application for processing. 


This work on a general purpose OS is typically split between two levels of scheduling; 1) interrupt level in which minimal work is done but at highest critical priority and is run-to-completion (typically called the top-half processing), 2) the so-called bottom-half uses kernel facilities to execute both device driver logic and protocol processing in a manner that can be preempted and rate limited. Regardless, all this work is done at the OS privilege level and ultimately prepares data for application processing (pre-emptable), and is independently scheduled at lower privilege and priority. 

An application specific library OS stack sheds much of the above complexity, both shortening the path and eliminating the above privilege scheduling domains~\cite{arrakis,ix,ebbrt}. It exploits short-cuts that allow run-to-completion execution of all the logic, including application processing in response to detecting device activity. 

\subsubsection{Application Processing}
\label{sec:workflow:appproc}
As illustrated in figure~\ref{fig:timeline}, during application processing, OS logic may be interleaved.  This work roughly falls into two categories, synchronous work done in service of this application request (page-faults, system calls, etc) and asynchronous work not having to do with this request (OS background work, processing of other requests or processes). Library OS's can often avoid interleaving asynchronous work, unrelated to the request handling,  and thus minimize jitter and improve IPC. 
\subsubsection{OS Reply Processing}
\label{sec:workflow:osrepproc}

At some point during application processing, a reply is generated and submitted to the OS for transmission. This can be handled in an asynchronous fashion depending on the OS semantics;  the OS can initiate protocol processing and device transmission in parallel with the remaining application logic (eg. book keeping, cleanup and preparation for the next request). This overlap reveals a potential opportunity for performance-energy trade-off. Specifically, it is possible given a particular packet arrival rate that slowing down causes the remaining application work to coincide with the time for the next request to arrive in both a closed and open loop setting, therefore keeping the processor busy. As such it may be possible that trade-offs in sleep state latency, interrupt overheads and polling leads to better performance at lower energy consumption.  


\subsubsection{Idle Policy}
\label{sec:workflow:idlepolicy}

If all processing is complete, no traffic is pending and aggressive polling is not in use, the OS can use a policy that selects a hardware sleep state, such as Intel C-states~\cite{intel_manual}, to halt the core. Various policies around optimizing them have been studied as well~\cite{dynsleep,dreamweaver,slowdownorsleep}. Each sleep state has an associated reduction in static power consumption. In the extreme, the deepest sleep states can flush micro-architectural state such as caches and power down these structures. However, each sleep state also imposes a progressively larger wake-up latency and potential impact on execution efficiency given the possible flushing of state~\cite{7425206}. 


There is clearly a relationship between the Idle Policy and Request Detection processing.  For a general purpose OS the normative assumption is both are interrupt driven.  Where an inter-dependency between the halt and interrupt mechanisms of the processor is exploited. In this study, we allow Linux's scheduler and default idle policy to decide if a core should be halted and to what state. This policy exploits various statistics to estimate how long the core is likely to be idle. It takes into account an estimate of when the next interrupt will likely occur from any source. This is a subtle implementation that interacts across many layers of the OS software, including the device driver. The idle driver framework also includes code provided by processor manufacturers to evaluate latency penalties and suggested minimum residency times. This allows us to see the impacts of making informed decisions regarding sleep states.  

In contrast, the library OS explores two simple policies: 1) when there is no work to process on a core, the processor is put into the deepest c-state (C7), thus ignoring any trade-offs in use of other sleep states\footnote{We have explored shallower c-states but focused on C7 as it had the most energy savings with minimal performance degradation.} so that we can focus on the interaction of slowing down the processor and adjusting interrupt delays with the use of a fixed deep sleep, and 2) using an aggressive poll loop on all cores to check for IO events such that the processor is always kept busy and no idle policy is used. 


\section{Performance and Energy}
\label{sec:slowdown}
In this section, we discuss the decomposition of figure~\ref{fig:timeline} timeline relating to slowing down the processor, delaying requests, and OS specialization for network processing.

\subsection{Interactions with Slowing Down the processor}
\label{sec:workflow:dvfs}
The use of DVFS in a processor allows software to adjust the energy consumption of CMOS based logic while trading off instruction execution speed. As noted in~\cite{slowdownorsleep, 10.1109/40.888701, pacingtoidle, udpm}, static or leakage energy consumption (i.e. caches, TLBs) is not particularly affected by DVFS but induces a base cost for keeping a fixed core architecture active.\footnote{Opposed to big-little or re-configurable core architectures.} An implication is that workloads which primarily use memory operations will suffer fewer performance penalties induced by a slowed processor while gaining energy saving benefits.

Our study explores how "slowing down" processing via DVFS interacts with the processing of network driven software stacks and the resultant energy and performance realized. We view DVFS as a speed control setting that can dilate CPU processing components of the request timeline in exchange for reduction in energy consumption.

From this perspective, the three obvious components that can be affected are OS Request, Application and OS Reply processing. For any given OS, there will be a hot-path instruction sequence that will be commonly exercised to process each request packet. The OS implementation will determine the type of instructions that will comprise of this path for a particular workload. As such, at the fastest DVFS setting there will be a characteristic mean number of cycles that will be required and thus an instructions-per-cycle (IPC) efficiency realized. It is important to note that better IPC does not necessarily imply better or worse performance or energy. What matters more is the amount of application work done per energy spent; as shown in~\cref{sec:mcdsilo:ipc}, a more efficient implementation that uses less instructions, though with worse IPC, can still result in better performance and energy efficiency.




\begin{table}[t]
\centering
\begin{tabular}{l|c|c|c}
  Name & Scenarios & Nature & CPU\\
  \hline
  NetPIPE & {\small 64B,8KB,64KB,512KB} & CL & Low\\ \hline
  NodeJS & na & CL & High \\ \hline
  Memcached & 200K, 400K, 600K & OL & Low \\ \hline
  Memcached-Silo & 50K, 100K, 200K & OL & High \\ 
\end{tabular}
\caption{Workload configurations.
The column {\em Nature} indicates open (OL) -versus-closed (CL) loop nature
and {\em CPU} indicates application work demand.}
\label{table:wrkcfgs}
\vspace{-.33in}	
\end{table}
\subsection{Interactions with Delaying Interrupts}
\label{sec:workflow:itr}
As observed in \cite{mootaz, udpm}, the "latency-slack" between the mean time to service a request and the SLA target of an application creates an opportunity for energy-performance trade-offs. Further, these works suggest creating energy management policies that delaying request processing in order to interact with DVFS and c-states in latency sensitive workloads to yield useful trade-offs. Both these works build an application specific prediction model of the time that is required to service requests and take a specification of the required tail latency SLA target. The  controller strategies exploit a combination of delaying processing and slowing down to find an optimal setting that ensures acceptable tail latencies while reducing the energy consumption.  The intuition is that there are advantages to using batching to consolidate idle time, therefore lengthening the time that the processor is in deeper sleep states and additional energy savings from slowing down processing given the latency slack.

A common feature of modern high speed NICs is the ability to delay the delivery of interrupt when an event such as packet arrival or transmission completion occurs. By manipulating this setting, software can limit the minimum time between interrupts or in other words the maximum rate at which the NIC events can interrupt the processor. The NIC used in this study exposes this mechanism via an Interrupt Throttling (ITR) setting~\cite{intelethtool}. Software uses the ITR register to configure a delay in 2$\mu$s increments.  If the spacing of events, such as packet reception, is less than  $2{\mu}s \times ITR$ the NIC will delay assertion.  If on the other hand events are sufficiently separated an interrupt will be asserted immediately.
By default the Linux device driver attempts to automatically set this interrupt delay value to reduce interrupt overheads.  We disable this feature and manually control its value to explore the impact of delaying interrupts on performance and energy. Delaying interrupts introduces an additional control that can interact with OS code and packetized payloads induced by Message Transmission Unit (MTU) constraints. If a request requires several MTU's then delaying interrupts can help reduce the interrupt processing overheads. Similarly, this can also induce prolonged quiescence periods in which processor idle policies can take advantage of.




\subsection{Interaction with Specializing OS Paths} 
\label{sec:workflow:osspec}
In this study, a baremetal library OS is used to reveal the value of OS path specialization against a general purpose OS, we believe this will help further motivate the adaption of specialization even in general purpose systems. OS specialization for a single application means that in figure~\ref{fig:timeline}, all of the Request Servicing in the timeline is affected. As the system no longer needs to support other processes and multiplex different devices, the entire software can be dedicated towards one use-case, furthermore, more application work can thus be done per instruction. In the case of a service oriented workload that has significant application work, such that the fraction of the instructions composed by OS network processing is small, there is a potential for improved performance and energy. Customized OS paths can both reduce the time spent in the OS processing and improve the application code IPC (\cref{fig:mcdsilo_detail}) due to reduction in architectural hazards associated with interrupts, protection domain crossing, etc. This time reduction can also increase the utility of using DVFS and delaying interrupts to find optimal settings for different workloads. 

Specialization also enables exploring alternate policies such as removing the Request Detection, Quiescent, and Idle Policy categories from figure~\ref{fig:timeline} altogether to keep the CPU always busy with a polling loop. This alternative policy simplifies the complex control problem of managing interrupt delays and sleep states usage, in addition, given that polling is a CPU operation and will interact with processor speed settings, it is thus possible that with efficient OS paths and a slowed processor via DVFS, polling can be used to find a more energy efficienct way of supporting different workloads (\cref{sec:closed_loop:poll} \cref{sec:mcd:poll}).

\section{Experiment Setup}
\label{sec:exp_setup}
\subsection{Hardware Platform}
Our experimental cluster consists of seven nodes,
each having 16-core processors of either 
Intel(R) Xeon(R) CPU E5-2690 @2.90GHz
or Intel(R) Xeon(R) CPU E5-2650 @2.60GHz type.
All processors have Intel 82599ES 10-Gigabit SFI/SFP+ NICs,
and are configured with a mix of 126 GB and 250 GB RAM.
The node used to boot into the baremetal library OS, EbbRT, and Linux uses a Intel(R) Xeon(R) CPU E5-2690 @2.90GHz processor with 126 GB of RAM. While the hardware used in this study are not modern, the two mechanisms used are still commonly supported~\cite{mellanoxsinterrupt, armdvfs, 10.1145/3184899}.

We ensured hardware hosting Linux and the library OS
are setup in a similar way by carefully configure IA-32 Architectural MSRs and processor specific MSRs
(see Tables 35-2 and 35-18 in ~\cite{intel_msr})
as well as NIC features:
direct-cache injection (DCA) disabled,
receive-side scaling (RSS) enabled
(to distribute packets for multi-core processing),
and hardware checksum offloading enabled.
We also match the values of
the number of NIC transmit and receive descriptors
and write-back thresholds for packet transmissions.
Additionally, to minimize system noise, hyperthreads and TurboBoost are disabled on all processors. While prior studies have included TurboBoost in performance-energy studies~\cite{udpm, pacingtoidle, PerAppPower, ixcp}, there have also been reports of energy anomalies when used with different sleep states~\cite{slowdownorsleep}. 

\subsection{OS Software}
\label{sec:OS}

\subsubsection{Linux}
\label{sec:OS_linux}
We build a set of application-specific Linux \textit{appliances} for the  four workloads listed in \cref{table:wrkcfgs}. These appliances are specially constructed to run a RAM-based filesystem and contain only a small set of system libraries and kernel modules required to run their constituent applications. We construct these appliances from a base Debian 10.4 distribution and use a custom 5.5.17 kernel which we built using a modified configuration file created for supporting high performance; following suggestions from previous work that studied Linux core operation costs~\cite{linux-core-ops}. To avoid scheduling overheads and noise, we pin all applications to physical cores. In addition, we disable Linux ~\textit{irqbalance} and affinitize packet receive interrupts to their respective cores.

\subsubsection{Library OS}
\label{sec:OS_libos}

We ported EbbRT~\cite{ebbrt}, an open sourced library OS, to run baremetal by developing a device driver for the Intel 82599 NIC~\cite{82599}. EbbRT is used as a platform for exploring specialization of OS paths and our findings are extensible to other systems since EbbRT shares similar structural properties (such as run-to-completion, event-driven execution model, single execution domain, and compile-time optimization) with other high performance OSes and systems developed for accelerating network workloads~\cite{ix, arrakis, zygos, shenango, rumpkernel, aliraza, unikernels, scalingmcdfacebook, arachne, mtcp, sandstorm, affinityaccept, flexnic, mica, seda}.

EbbRT consists of specialized components written in C++\footnote{ All components are multi-core functional and optimized to aggressively use per-core memory and fine grain locking.}, with a NIC driver, a custom TCP/IP stack, virtual and physical memory allocators\footnote{Memory allocators make aggressive use of large pages and pinned memory to avoid page-faults.}, and a generic I/O buffer\footnote{I/O buffers are designed to enable zero-copy application data processing.} It is packaged as a library of configurable modules and \textit{gcc-5.3.0}-based tool-chain targeting the base components of the OS. 

Applications are ported to it by configuring the necessary OS components and compiling the application source along with any dependent libraries using this tool-chain. This generates a single application-specific binary that is compile and link-time optimized with the OS code. Our port enables application-specific binaries to boot directly on our hardware platform. Once booted, OS and application code is executed under a single supervisor privilege domain.

Given the design and implementation for single-application, non-preemptive processing via an optimized OS and application binary, library OS components can avoid many checks and streamline execution, ranging from interrupt dispatch to application logic. The NIC device driver totals over 3000 lines of code and interfaces with EbbRT's multi-core TCP/IP network stack\footnote{The device driver programs the NIC using per-cpu queues and interrupts, maintaining the affinity of TCP connections to their respective cores.}. EbbRT provides an interface for statically setting interrupt delay values. We use this interface in our study as we sweep across interrupt delay values. The NIC driver also exposes a configurable constant (set to 64 for all our experiments) that is used to control how many packets can be processed in a single interrupt invocation before returning to the event-loop of the core on which the interrupt was processed. This behaviour, in turn, introduces a simple bounded per-cpu device-level poll.

\subsubsection{NIC polling without sleep}
The simple \\
run-to-completion, and lightweight event-driven execution model of EbbRT allows us to also explore the performance-energy trade-offs of slowing down the processor in the context of a polling loop for packet processing. We use standard techniques to auto clear hardware interrupts and enable a tight polling loop. The loop checks a in-memory data structure in which the NIC updates whenever new packet descriptors are to ready be processed. Due to this tight loop, EbbRT will never halt the processor and thus will not use any sleep states.


\begin{figure}
\centering
\includegraphics[width=1.0\columnwidth]{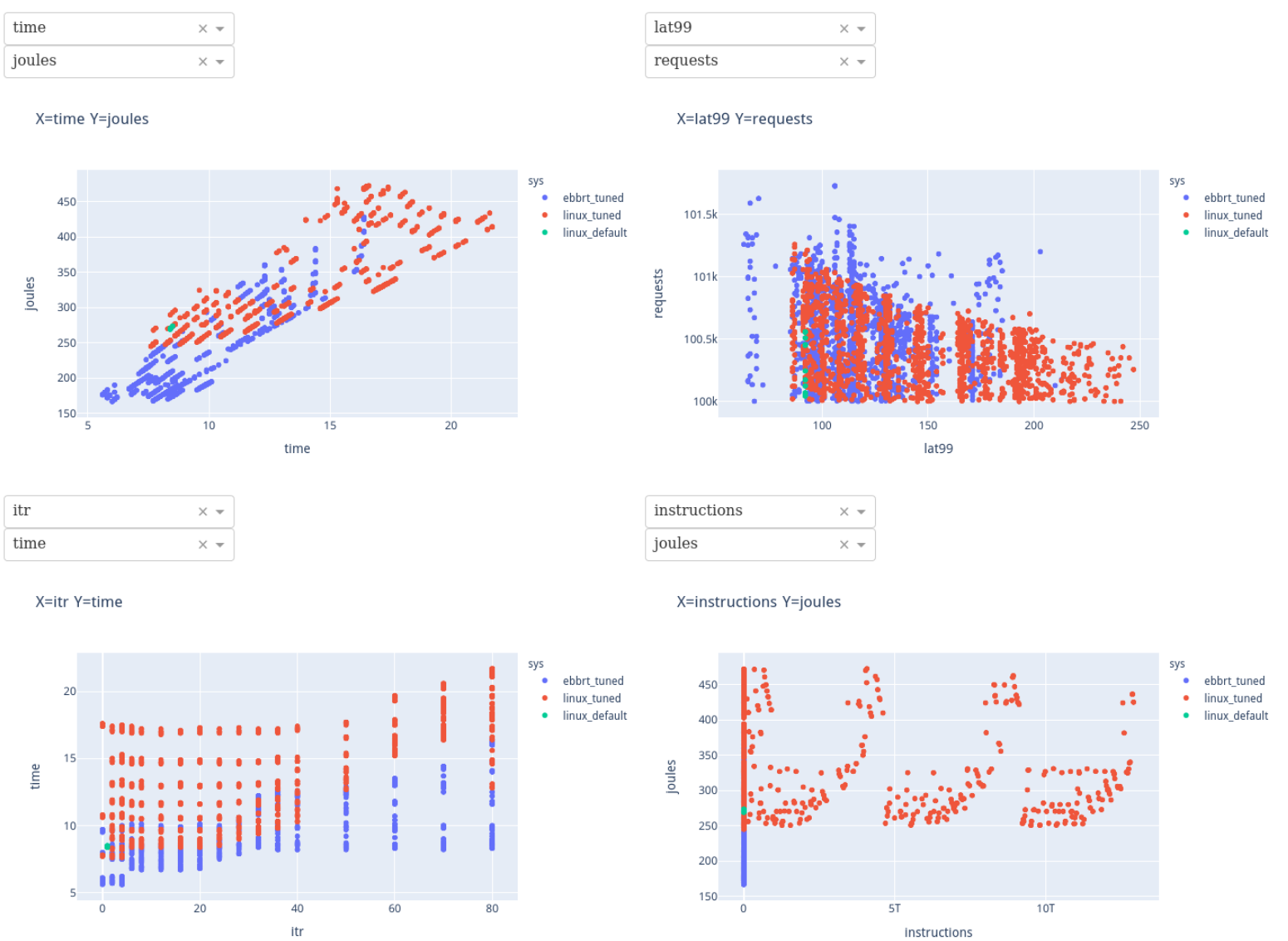}
\vspace{-0.15in}
\caption[]{\small Small example of web visualization tool.}
\label{fig:webapp}
\vspace{-0.15in}
\end{figure}

\subsection{Per-Interrupt Log Collection and Visualization}
In order to better understand the interactions of interrupt-delay and DVFS under a workload, we instrument fine-grained per-interrupt log collection in both Linux and EbbRT's network device driver. We collect the following information in the NIC's interrupt handler code: received and transmitted bytes, received and transmitted descriptors, sleep state statistics, and the current timestamp (via \texttt{rdtsc} instruction). In addition, we instrument per-core performance monitoring counters (PMCs) to collect a set of hardware statistics after every millisecond\footnote{The millisecond gap is due to sampling granularity of RAPL.} of elapsed time: instructions, cycles, last-level cache misses, and standard RAPL hardware registers on Intel processors to read per package energy values~\cite{intel_rapl} as it has been experimentally validated for accuracy in previous works~\cite{rapl, rapl2015, rapl2018, weaver_rapl}. While we have validated results against rack-level energy measures (slowing DVFS and interrupt delay resulted in rack level energy savings), we use RAPL instead because the granularity of the rack level measurements (on the order of seconds) made it difficult to attribute detailed energy use to specific system events.


Given these collected log traces, we built a web visualization tool using Dash~\cite{dash} that enables a user to dynamically examine system behaviour across a wide range of configurable settings, for example, figure~\ref{fig:webapp} shows how one can view the data at different dimensions (via dropdown boxes) of interrupt delay value, processor frequency, instructions, cycles, time, etc. With a fine-grained log trace, we also used the tool to zoom in on specific events that transpired in-between hardware interrupts to 1) gain better insights at a fine-grained manner, and 2) to generalize these insights into broader findings as will be discussed in \cref{sec:exp}. Having this tool gave us the ability compare and contrast different OS behaviors and was also immeasurably helpful to visually understand the structure in the data.



\section{Experimental Analysis}
\label{sec:exp}
The main methodology we used in our study is that of manually setting DVFS and interrupt delay values to all possible values as an exhaustive search to find optimal performance and energy trade-off points for both Linux and the library OS, EbbRT.
We refer to these OS setups as \textit{Linux-tuned} and \textit{LibOS-tuned}, respectively, in the figures below. To better understand the degree of trade-offs in Linux, we also ran experiments on a base configuration which we refer to as \textit{Linux-default}; in this mode, Linux's interrupt delay and processor speed are both controlled dynamically by its built-in policies~\cite{cpufreq_governor,intelitr}.
We also explored a version of slowing down the processor by replacing network interrupts with a polling loop (whereby no sleep states are used) in EbbRT; we refer to this setup as \textit{LibOS-poll} in the figures.

Figures~\ref{fig:closed_loop_overview},~\ref{fig:mcd_overview},~\ref{fig:mcdsilo_overview} shows overviews of all the experimental runs gathered across the different applications and their respective loads as listed in table~\ref{table:wrkcfgs}. Each data point represents a single experimental run and each experiment is repeated ten times for stability.
For each workload, we break down the trade-offs observed by manually setting DVFS and interrupt delay in both OSes into 
measurements of performance (e.g. time for closed-loop workloads and 99\% tail latency for open-loop workloads) and measurements of energy use.
In order to reason about these trade-offs, we use two graphical mechanisms to highlight the differences:
\begin{enumerate}
    \item The \textit{size} of each point represents the degree with which interrupt delay is used; the {\larger[1]\textit{larger}} the size, the more interrupt delay value is \textit{increased} while the \textit{smaller} the size the more it is \textit{decreased} (e.g. faster IO interrupts).
    \item The \textit{color gradient} of each point represents the degree of slowing down processor speeds; the \textbf{darker} the color the more the processor has been \textit{slowed} (less energy use) and vice-versa \textit{faster} when the color is \textit{lighter} (more energy use).
\end{enumerate}
Finally, for each of the system configurations studied, the configuration that yields the best performance and lowest energy is indicated with {\larger[4]\textbf{+}} and {\larger[4]\textbf{x}} respectively.
\subsection{Closed Loop Workloads}
\label{sec:closed_loop}
\begin{figure*}
\centering
\includegraphics[width=1\textwidth]{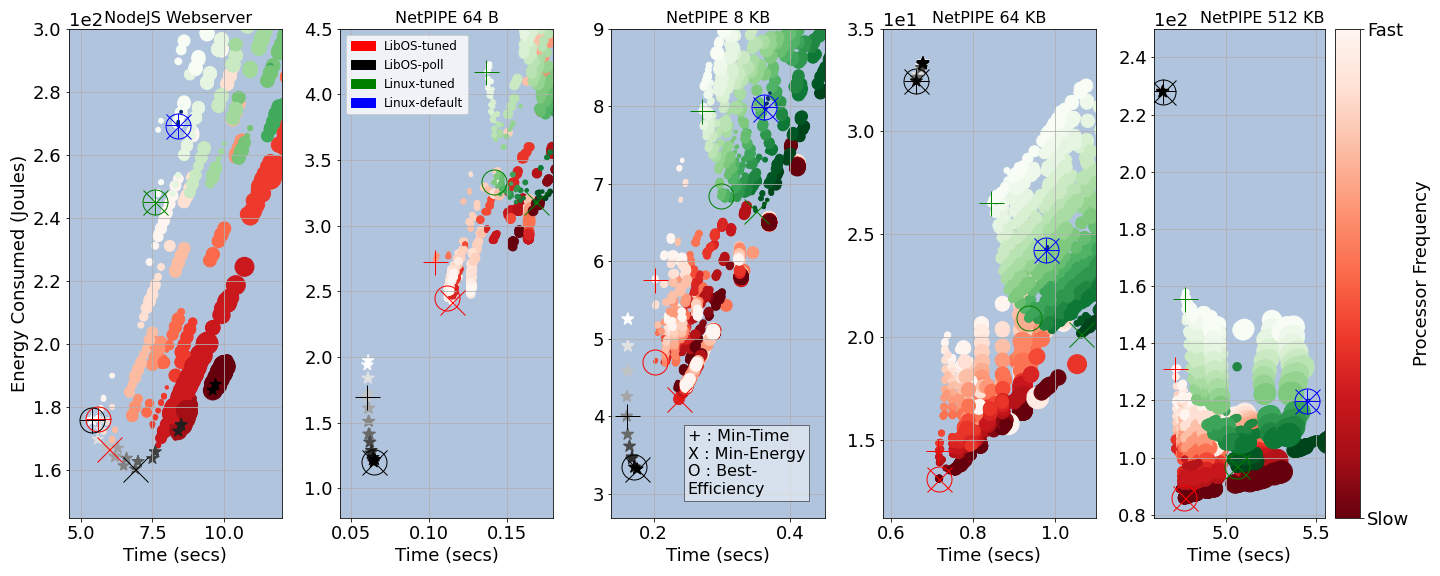}
\vspace*{-10mm}
\caption[]
{Overview of closed loop experiments (\textbf{X and Y axis are scaled differently to expose structure}). LARGER dots uses SLOW interrupt delay;
Other visual cues: A \textbf{darker} color indicates \textbf{slow} processor frequency; \textbf{X} indicates lowest energy consumption; \textbf{+} indicates lowest time spent; \textbf{O} indicate best energy efficiency. Polling (grey-black datapoints) results in best energy efficiency for NodeJS and NetPIPE at 64 B, 8 KB message sizes. For interrupt-based configurations, a fast interrupt delay value is used to achieve best efficiency. At 64 KB and 512 KB for NetPIPE, best efficiency uses slow DVFS with a slow interrupt delay value that trades off speed and energy savings most efficiently. Polling largely inefficient at 64 KB and 512 KB.}
\label{fig:closed_loop_overview}
\end{figure*}

\begin{figure*}
\centering
\vspace*{-3mm}
\includegraphics[width=1\textwidth]{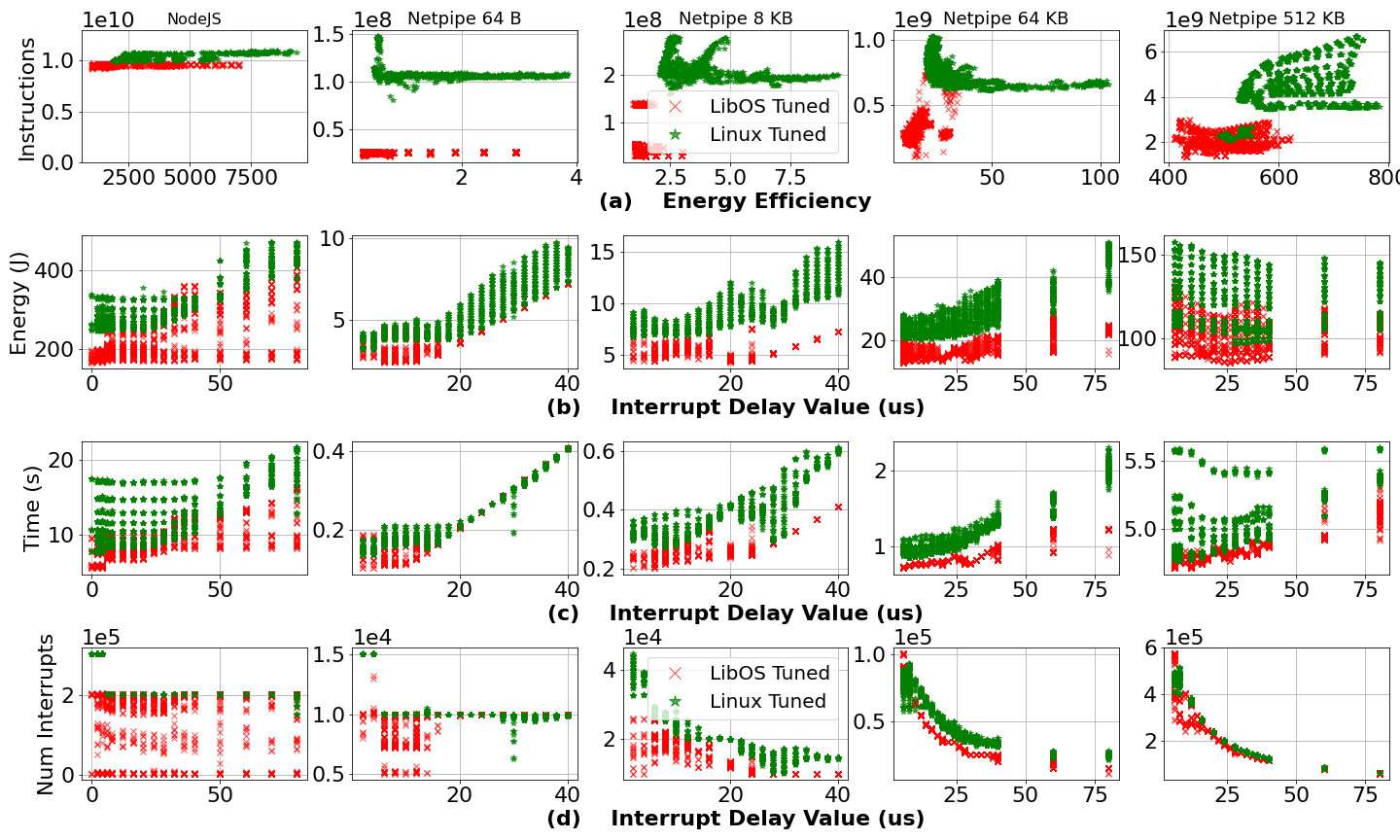}
\vspace*{-10mm}
\caption[]
{Detailed plots of some gathered statistics for the above closed loop experiments. (\textbf{X and Y axis are scaled differently to expose structure})
These plots only compare LibOS-tuned and Linux-tuned. All figures in a vertical slice is mapped to the same workload. Though (a) shows similar amount of instructions in both OSes, (b)(c) shows LibOS-tuned using both less energy and time to finish same workload, indicating efficiency of path specialization. In (d) for NodeJS and Netpipe 64 B, the big differences in interrupts is a result of slow-to-stay-busy effect as described in \cref{sec:closed_loop:slowtostaybusy}.
}
\label{fig:closed_loop_detail_1}
\end{figure*}


Figure~\ref{fig:closed_loop_overview} illustrates the set of closed-loop workloads that we study, all of which are run on a single core with a single connection.
Netpipe~\cite{snell1996netpipe} involves sending messages of identical size between two systems for a fixed number of iterations.
We run Netpipe in a symmetric configuration, whereby the client and server sides run the same software stack and are configured with the same performance parameters. This approach allows us to analyze Netpipe performance precisely, as it eliminates any potential ambiguity in the measurements that may arise from computational differences between the communicating client and server. While Netpipe isn't a realistic workload in the datacenter, it allows us to explore different message sizes, opening up the scope of how the DVFS and interrupt delay affects time and energy.
Linux runs NetPIPE-3.7.1 while the library OS uses a custom version ported to its interfaces.
We fix the iteration count at 5000 and show results for a range of message sizes \footnote{We found that the 10 GB link is close to saturation when a message of size greater 700 KB is exchanged.}.
As message size increases, the workload becomes more network bound; Linux suffers an additional memory copy from kernel to userspace compared to the libOS.

NodeJS~\cite{nodejs} consists of a JavaScript HTTP Webserver running inside a nodejs runtime. A single client running the \textit{wrk-4.0.2}~\cite{wrk} benchmark\footnote{We modified \textit{wrk} to place a fixed request load of 100K.} sends requests to the server for a fixed period of time. The server responds to each request with a small static payload of size 148 bytes.
Linux runs nodejs-0.10.46, and the library OS runs the same version ported to support baremetal nodejs by providing OS interfaces that link with the V8~\cite{v8} JavaScript engine and libuv~\cite{libuv}. 


Given the nature of closed loop workloads, one would ideally minimize both time and energy. Therefore we use a single value, the product of time and energy, as a measure of energy efficiency for comparison between the two OSes.

\subsubsection{Reducing time to save energy for small payloads}
\label{sec:closed_loop:speedup_small_payload}
One mechanism to reduce time across all the closed loop workloads is to always use a low interrupt delay value as shown in figure~\ref{fig:closed_loop_detail_1}(c). With small payloads, reducing time equates to reducing energy (see figure~\ref{fig:closed_loop_detail_1}(b)) as well. For nodejs and netpipe 64B, setting a low interrupt delay (2$\mu$s) resulted in efficiency improvements by 2X in Linux-tuned and a further 85\% in LibOS-tuned. This is due to the lightweight nature of the payloads and in this case, simply getting the work done fast leads to best energy efficiency.  

\subsubsection{Effects of interrupt delay induced batching on performance and energy}
\label{sec:closed_loop:speedup}
Polling (dark datapoints) results in best energy efficiency for NodeJS and NetPIPE at 64 B, 8 KB message sizes and for interrupt-based configurations, a fast interrupt delay value is used. In contrast, as netpipe payload sizes increased to 8KB, 64KB, and 512KB, the interrupt delay value that yielded best energy efficiency also became slower (up to 28$\mu$s at 512 KB). A 10 GbE NIC, assuming no network jitter and switching cost, can transmit at an optimal rate of 1250 bytes/$\mu$s. \textit{Therefore, the interrupt delay value can be used to effectively determine how much payload the software should process in a fixed quantum}. With larger message sizes, one can imagine portions of its payload being transmitted over the wire and processed by software asynchronously. The interrupt delay value that yields best efficiency is indicating a "sweet spot" with which the software should pace packet processing and save energy by sleeping during the its quiescent periods.

By adjusting interrupt delay values in accordance with particular payload sizes, Linux-tuned exhibits improved energy efficiency over Linux-default up to 80\%. Due to specialization of the library OS, figure~\ref{fig:closed_loop_detail_1}(a) shows that it always uses fewest instructions, even in computationally heavy workloads such as nodejs. This efficiency, coupled with a custom interrupt delay, enables LibOS-tuned to improve its energy efficiency over Linux-tuned by another 2X.


\subsubsection{Trade-offs in library OS polling}
\label{sec:closed_loop:poll}
We compare the performance and energy trade-offs between slowing down the processor while the library OS is in a polling loop (LibOS-poll) and slowing down both processor and interrupt delay (LibOS-tuned).
For nodejs, LibOS-poll only results in a 4\% better energy efficiency than that of LibOS-tuned, primarily due to nodejs runtime already using an application-level poll to check for new packets.

The difference in energy efficiency for LibOS-poll is quite dramatic as message size increases in netpipe. With 64B and 8KB message sizes, polling improves this efficiency by 1.6X and 3X respectively over LibOS-tuned (11X over Linux-default). This is because for smaller payload sizes, getting the work done fastest results in the lowest energy use and mirrors the explanation in \cref{sec:closed_loop:speedup_small_payload}. At 64KB and 512 KB, the workload becomes more network bound and as a result, polling results in worse energy efficiency by up to 2X compared to LibOS-tuned. At these larger message sizes, polling only reduced time by around 10\% while energy consumption increased over 2X than interrupt-driven LibOS-tuned, which is indicative that packets spent more time on the wire than in software. This phenomena suggests the importance of a hybrid strategy that switches between poll and interrupt-driven OS policies as payload size changes.

\subsubsection{Overlapping work with IO}
\label{sec:closed_loop:slowtostaybusy}
In both nodejs and netpipe with message size 64B, we find another interesting effect. In figure~\ref{fig:closed_loop_detail_1}(d), for all the various interrupt delay values, the total number of interrupts can be lowered by 90\%. Upon closer examination, we find slowed DVFS caused this decrease in number of interrupts.

The reason for this behavior in the library OS is described briefly in \cref{sec:workflow:osrepproc}:
The physical transmission of OS reply packets by the network driver can occur asynchronously with the unwinding of the stack back to the nodejs application and then back down to the network receive function to check for new packets.
The slowing down of the processor causes this unwind path to lengthen, potentially increasing the probability that new packets have already arrived ready to be processed by the time it reaches the network receive function.
Therefore, the software is able to skip one or more hardware interrupts (fired on packet receive) in order to effectively \textit{slow-to-stay-busy} and process this new reply packet. This scenario only occurs in the library OS due to its run-to-completion nature and suggests that, for a structurally different OS, other energy saving strategies can be explored.





\subsection{Open Loop Workloads}

\subsubsection{Memcached}
\label{sec:mcd}

\begin{figure*}
\centering
\includegraphics[width=1\textwidth]{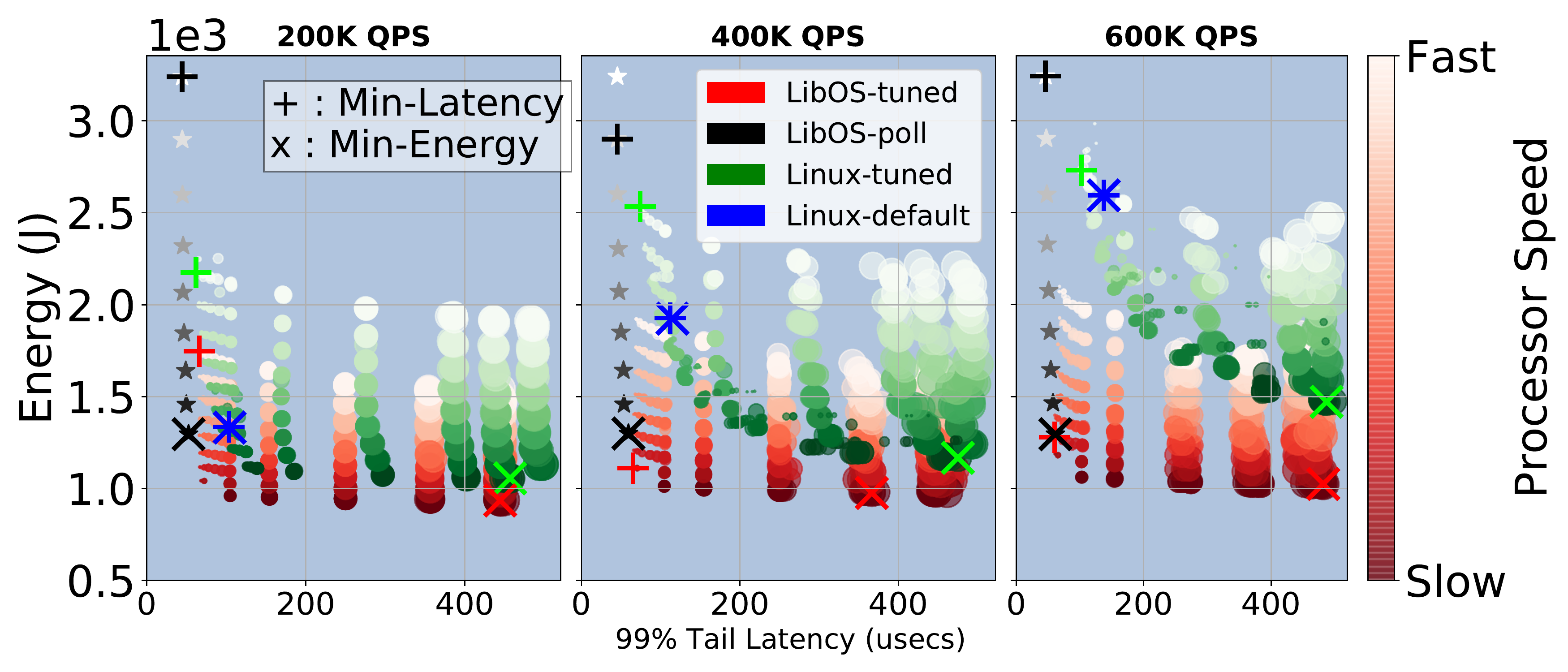}
\vspace{-.4in}
\caption[]
{ 
Overview of memcached experiments. \textbf{x} indicates lowest energy. \textbf{+} indicates lowest latency. In contrast to Linux, the vertical nature of LibOS-tuned shows its efficiency and instructions less impacted by slowed DVFS. Distance from Min-Latency to Min-Energy in contrast to Linux-default shows potential in exploring trade-offs through coordinated use of both mechanisms.
}
\label{fig:mcd_overview}
\vspace*{-.2in}
\end{figure*}

\begin{figure}
\includegraphics[width=0.47\textwidth]{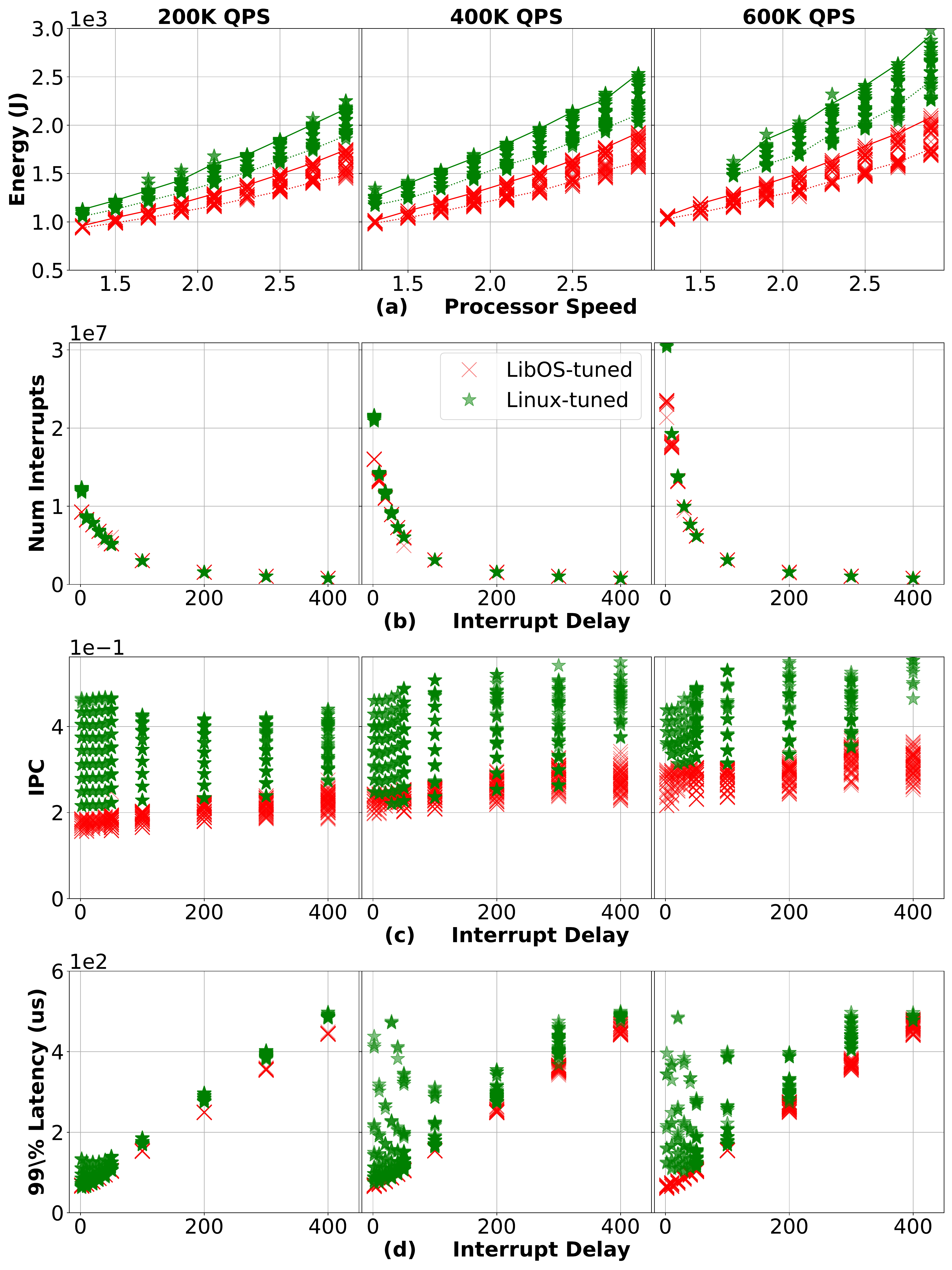}
\vspace{-.22in}
\caption[]
{
Some detailed plots between LibOS-tuned and Linux-tuned in memcached. Each vertical figure is mapped to the same QPS. In (a), bold lines connect points that use fastest interrupt delay and dashed lines use slowest interrupt delay; this shows additional energy savings induced by batching. In (d), one can see stability of tail latency values of LibOS-tuned compared to Linux-tuned.
}
\label{fig:mcd_detail_1}
\end{figure}

\begin{figure}
\includegraphics[width=0.5\textwidth]{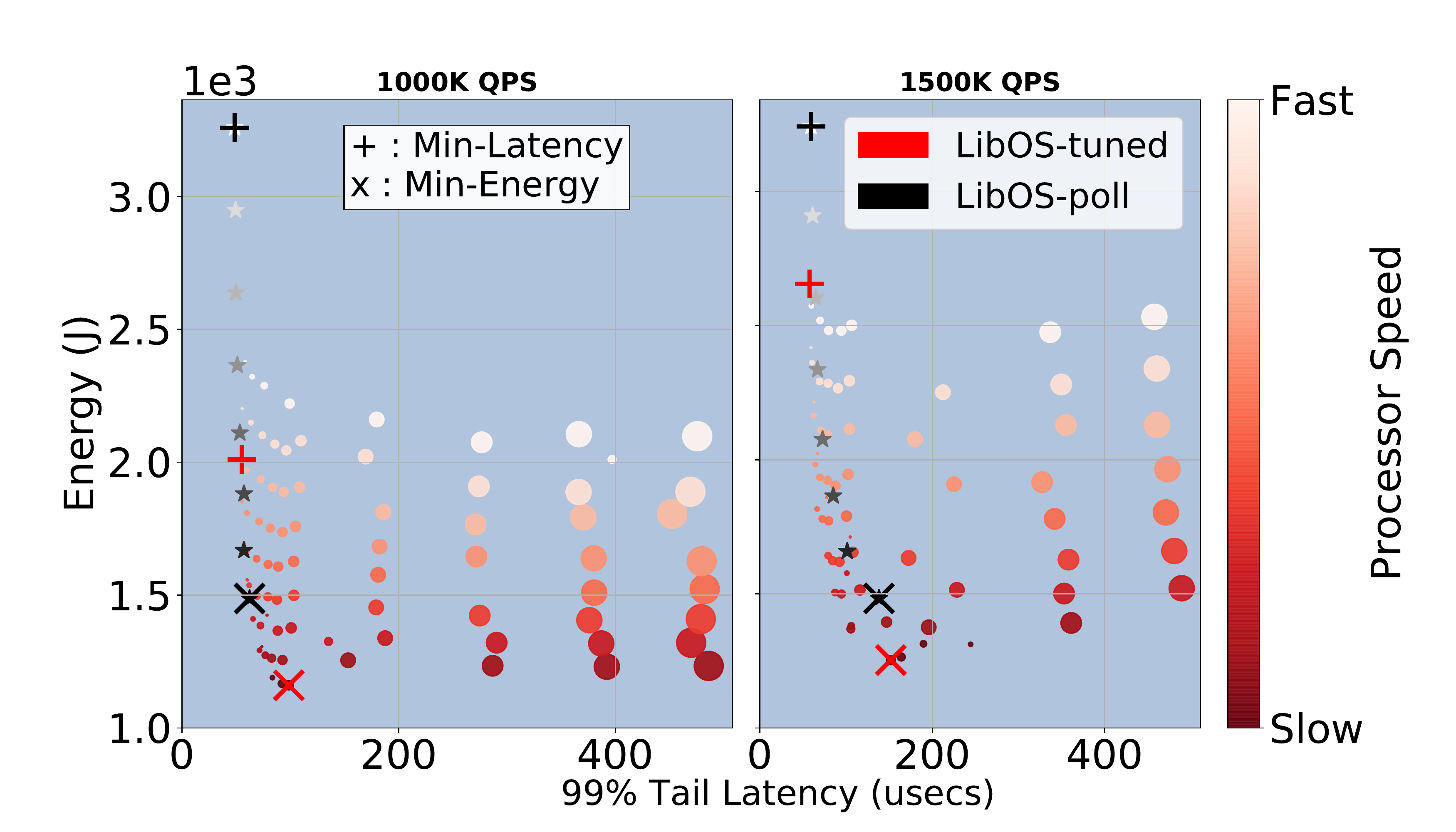}
\vspace{-.28in}
\caption[]
{
Overview of memcached experiments at 1000K and 1500K QPS. Vertical nature of library OS largely holds even at higher QPS loads and polling still competitive with interrupt-based in both tail latency and energy.
}
\label{fig:mcd_overview2}
\vspace*{-.2in}
\end{figure}







This is a multi-threaded open loop workload that runs on all 16 cores of a server node~\cite{mcd}. It consists of an unloaded client node running mutilate~\cite{mutilate}. This client (1) coordinates with five other mutilate agent nodes in order to generate requests to the server and (2) measures tail latency of all requests made. All five agent nodes are 16-core machines, whereby each core creates 16 connections, for a total of 1280 connections. This setup is able to saturate the single 16-core server\footnote{Mutilate is configured to pipeline up to four connections to further increase its request rate.}.

Linux runs memcached-1.6.6 and the library OS version uses a re-implemented version of memcached, written to EbbRT's interfaces, and supports the standard memcached binary protocol. We run a representative load from Facebook~\cite{workloadanalysisfacebook} (ETC) which represents the highest capacity deployment. It uses 20 to 70 byte keys and 1 byte to 1 KB values and contains 75\% GET requests.


\textbf{Impacts of Slowing Down in Different OSes Structures}: \\
\cref{fig:mcd_detail_1} shows that although LibOStuned has worse IPC than Linux-tuned across the three QPS loads; it used on average 2.5X fewer instructions than Linux, which implies a greater fraction of its instruction were spent getting the work done. Furthermore, given that memcached is not compute heavy, most of its instructions are therefore memory bound; this also lowers the effect of a slowed processor to increase tail latency. The vertical nature of the library OS in \cref{fig:mcd_overview} and \cref{fig:mcd_overview2} illustrates this behavior. Furthermore, this suggests the logic and data structures used by a specialized library OS results in instruction mixes that can take advantage of energy saving benefits of slowed DVFS without sacrificing performance.

Figure~\ref{fig:mcd_overview2} also shows that the library OS can support higher QPS loads than Linux due to its specialized paths. We see the opposite of this behavior in Linux where, at 600K QPS, it approaches 75\% of its peak QPS. There is also a clear trade-off between slowing down processor speeds and an increase in tail latency (higher latency points have darker gradient color). In figure~\ref{fig:mcd_overview2}, memcached is scaled higher to 1500K QPS, which is 75\% of the peak QPS of library OS. At this QPS rate, we can begin to see similar trade-offs in both OSes. 

\textbf{Impact of Slowing Down Processors on Sleep States}: \\
Figure~\ref{fig:mcd_detail_1}(a) shows that as a processor slows down, the energy savings from slowing down interrupts also decrease. In this figure, bold lines indicate the mean energy use at fastest interrupt delay, while dotted lines indicate mean energy use at the slowest. We find that across the QPS loads and the two OSes, the average energy savings from slowing down interrupt delay at the \textit{slowest} processor speed is \SI{52}{\joule} while it is \SI{342}{\joule} at the \textit{fastest} processor speed. As discussed in \cref{sec:workflow:dvfs}, the effect of slowing down the processor results in the lengthening of the application and OS work. Therefore, this potentially reduces the energy savings that are brought about by taking advantage of sleep states during prolonged idle periods. Furthermore, such slow-down is undesirable due to SLA requirements which result in stringent time budgets that requests must adhere to.

However, figure~\ref{fig:mcd_overview} shows that it is a combination of slow DVFS and interrupt delay that results in lowest energy use across both Linux and the library OS. Figure~\ref{fig:mcd_detail_1}(d) shows the direct effect of interrupt delay on tail latency. The benefit of slowing down interrupts consists of 1) lowering the number of interrupts fired, which also lowers instruction use and potentially promotes better packet coalescing (see figures~\ref{fig:mcd_detail_1}(b)(c)), and 2) ensuring a guaranteed period of quiescence such that the processor can take advantage of potentially deeper sleep states. However, these trade-offs will be different dependent on other factors such as an OS's packet processing efficiency and policies that govern the use of sleep states to maximize idle states. Moreover, the benefit of slowing down interrupts versus processor speed is subtle as the implications of slowing down the processor affects entire software stack whereas interrupt delay has a fixed impact.






\textbf{Benefits of Interrupt Delays on Performance}: \\
\label{sec:mcd:fastitr} 
Figure~\ref{fig:mcd_overview} also demonstrates the ability to use a static fast interrupt delay value in order to minimize tail latency (smaller dots). Linux-tuned improved its tail latency over Linux-default by 40\% at 200K QPS and 25\% at 600K QPS. As discussed in section~\ref{sec:workflow:hybridio}, a faster interrupt delay can induce a form of polling with Linux's NAPI policy by constantly waking up the processor to do the OS and application work. This induced behavior also increases energy use by 38\% at 200K QPS and 5\% at 600K QPS which represents another space in the energy-performance trade-off of memcached. While prior research have used static setting of a fast interrupt delay value for experimental stability~\cite{arrakis, shenango}, we are the first to show its energy implications.

\textbf{Polling Can be Energy Efficient}: 
\label{sec:mcd:poll}
Similar to Netpipe with 64B messages (see figure~\ref{fig:closed_loop_overview}), figure~\ref{fig:mcd_overview} shows that using an OS poll for network-bound workloads with a small payload results in the best performance (tail latency) in memcached. Although memcached is a more complex workload than Netpipe with thousands of connections and requests pipelined and multiplexed on multiple cores, LibOS-poll running memcached can still be energy efficient through slowing down the processor. Using \textbf{X} point (Min-Energy) of LibOS-poll as reference, we find that at 200K QPS, LibOS-poll improves both 99 percentile tail latency by 27\% and energy by 35\% over LibOS-tuned at Min-Latency point. As the load increases, by comparing the Min-Energy points, we find that while LibOS-poll consumes 11\%-38\% more energy across the rest of QPS loads, its tail latency was 10\%-90\% better. Hence a library OS poll reveals an additional trade-off space for energy and performance in memcached.







\begin{figure*}
\centering
\includegraphics[width=1\textwidth]{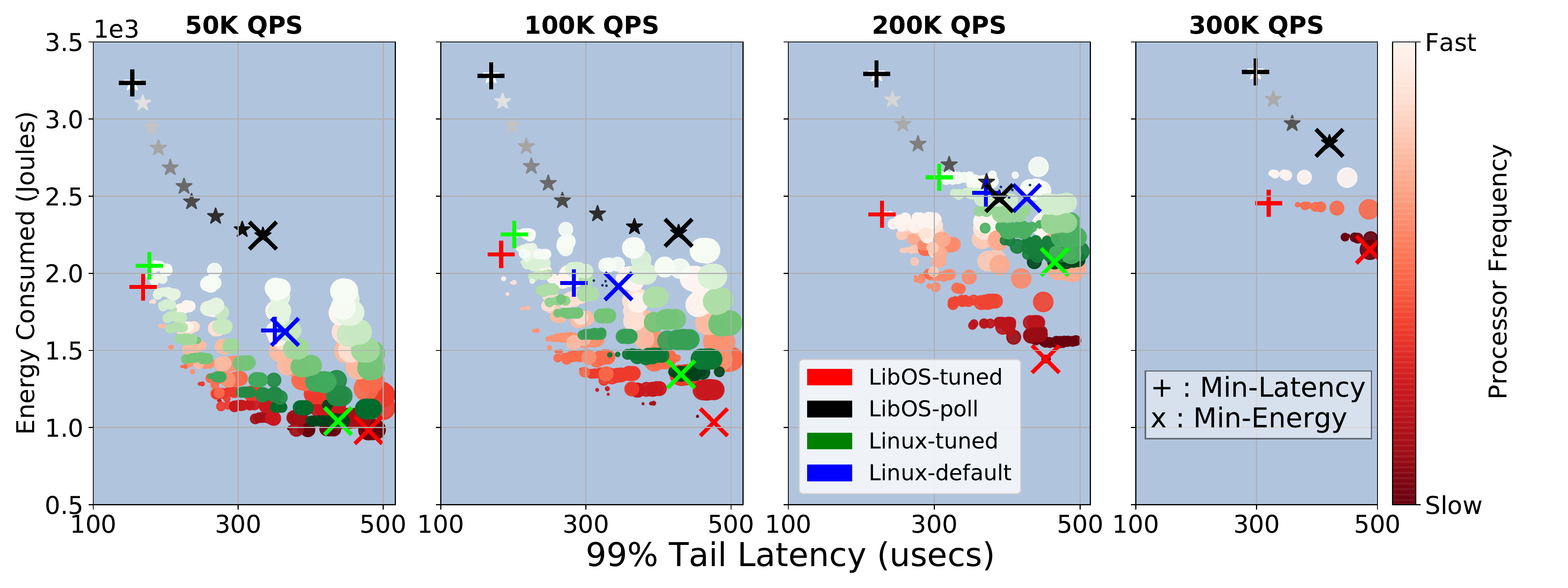}
\vspace*{-6mm}
\caption[]
{Overview of memcached-silo experiments across 50K, 100K, 200Km and 300K QPS. As application work gets larger, horizontal nature of points indicate slowed processor affects latency more, however, can still explore performance and energy benefits with DVFS and interrupt delay induced batching in both LibOS-tuned and Linux-tuned. The base tail latency of this workload is larger than memcached due to nature of TPC-C application work. Polling is largely energy inefficient compared to interrupt-based but is competitive in tail latency.}
\label{fig:mcdsilo_overview}
\end{figure*}

\begin{figure}
\includegraphics[width=0.5\textwidth]{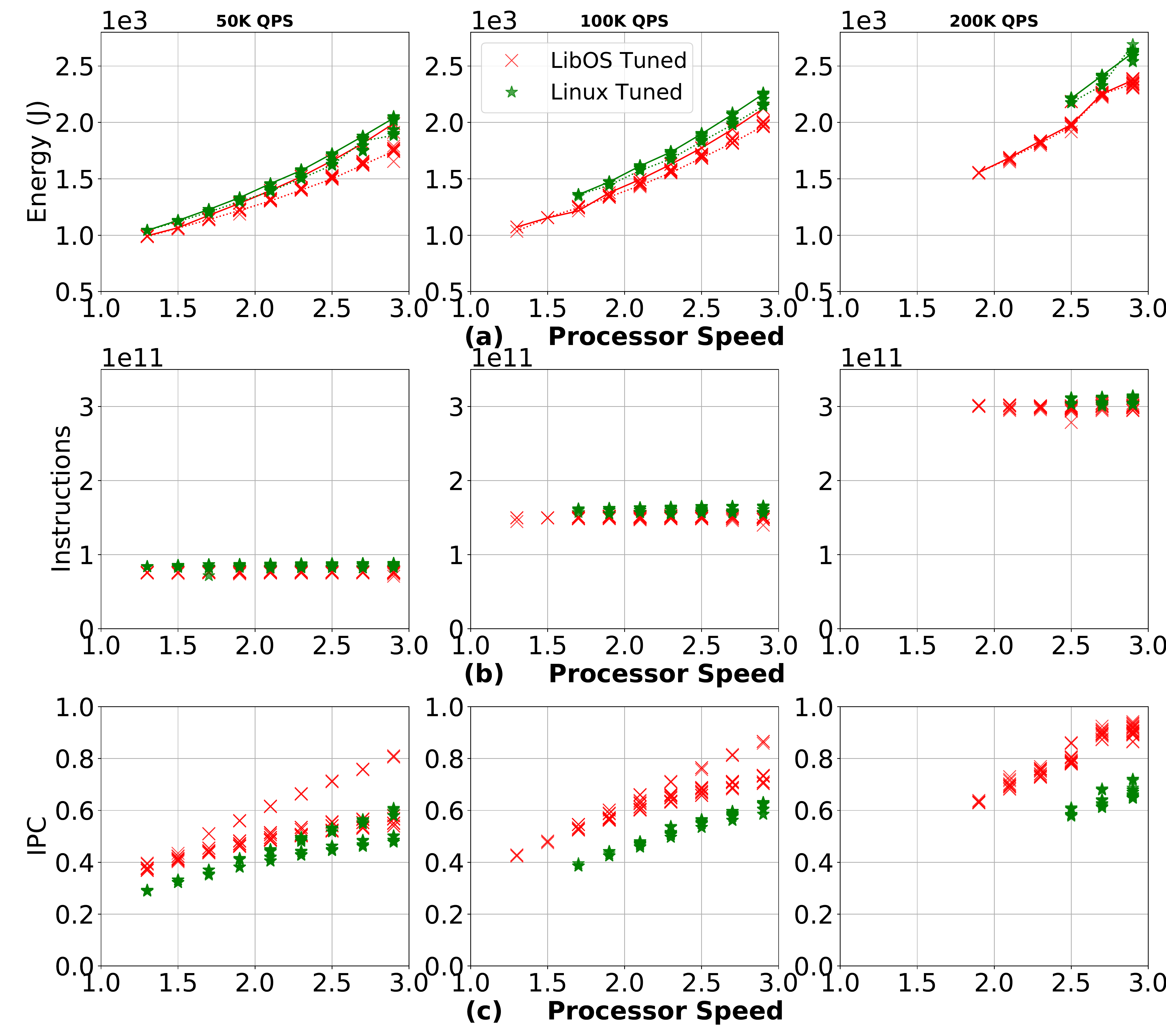}
\vspace*{-6mm}
\caption[]{Detailed plots of some gathered statistics of the above QPS loads. Compares LibOS-tuned and Linux-tuned. Each vertical figure is mapped to the same QPS. In (c), with an application heavy workload, LibOS-tuned actually achieves better IPC than Linux-tuned even though (b) shows similar amount of instructions; indicating more work done per instruction for a specialized OS.}
\label{fig:mcdsilo_detail}
\vspace*{-.1in}
\end{figure}

\subsubsection{Memcached-silo}
\label{sec:mcdsilo}

This is a workload built on top of the normal memcached protocol. It is more intensive both in computation and memory-use than regular memcached, as it is structured such that every memcached request triggers a corresponding set of TPC-C transaction processing logic on a in-memory database~\cite{silo}. We ported the memcached-silo implementation from~\cite{mcdsilo, zygos} to EbbRT. The workload mix and SLA constraints of memcached-silo follow from those used in the memcached experiments in \cref{sec:mcd}. However, given its heavier application nature, we only needed two 16-core client nodes at 16 connections per core to saturate a single 16-core memcached-silo server. 

\textbf{Slowing Down has Benefits Even in Computationally Heavy Workloads}:\\
\label{sec:mcdsilo:dvfstradeoff}
In contrast to memcached workload shown in figure~\ref{fig:mcd_overview}, figure~\ref{fig:mcdsilo_overview} shows that as the application work gets larger for each memcached request; the trade-offs of tail latency and energy become more discernible in both Linux and the library OS (note the color gradient darkens horizontally instead of vertically).
Surprisingly even in a computationally heavy workload under a stringent SLA, it is still possible to delay both processor and interrupts to further save energy; in figure~\ref{fig:mcdsilo_overview} at 200K QPS, Linux-tuned improved its tail latency by 21\% and energy by 20\% over Linux-default, further LibOS-tuned improved its tail latency by 34\% and energy by 44\% over Linux-tuned. Figure~\ref{fig:mcdsilo_detail}(a) shows that slowing the processor via DVFS results in the most energy savings and using interrupt delays on top leads to additional savings. However, in contrast to figure~\ref{fig:mcd_detail_1}(a), the effects of delayed interrupts are greatly diminished as the QPS load increases given the workload nature. One can see this effect in the 100K and 200K QPS experiments where the configuration that yielded Min-Energy for the library OS is represented by a significantly small dot, implying a a fast interrupt delay value. This observation harkens back to the implications of a slowed processor on the packet processing path (\cref{sec:workflow:dvfs}); furthermore, this effect is exacerbated as application logic gets heavier.

\textbf{Library OS IPC Efficiency}:
\label{sec:mcdsilo:ipc}
Even though figure~\ref{fig:mcdsilo_detail}(b) demonstrates similar number of instructions between Linux-tuned and the LibOS-tuned, the IPC measurement in figure~\ref{fig:mcdsilo_detail}(c) reveals that a specialized OS can execute instructions more efficiently even in an application bound workload. This IPC benefit not only leads to energy efficiency but also creates more slack for the library OS to take advantage of slowed down processor. Figure~\ref{fig:mcdsilo_detail}(a) shows that at faster processor speed values in the 50K and 100K QPS experiments, the library OS was still able to slow down interrupt delays to save more energy than Linux by 50\% even in a computationally heavy workloads.

\section{Related Work}
\label{sec:related}







Our work falls within a wider space of research on energy proportional computation in datacenters~\cite{energyproportion, warehouse-power, 268014}. Much of this research stems from the challenges of improving the performance of network-bound datacenter workloads like MapReduce~\cite{large-scale-mapreduce} and in-memory key-value stores~\cite{mica, zygos} while keeping energy consumption at bay. These challenges can be attributed to
the complex diurnal trends that are characteristic of datacenter-level utilization, whereby idle time is common and must be optimized for~\cite{hotpower2008, powernap, napsac} while simultaneously maintaining the ability to support high-utilization peaks and strict latency constraints ~\cite{Dynamo, SmoothOperator, oldi-pegasus, adrenaline, rubik, eurosys14, zygos, peafowl, 7425206, 10.1145/2830772.2830779, dreamweaver, dynsleep, udpm}. Our work examines both in-memory key-value stores and its modified version with a heavier processing component as well as closed loop applications. Our goal was to gain better insight into the systemic impacts of performance and energy when slowing down network workloads using the two hardware mechanisms of DVFS and ITR delay together.

There is a wide range of work that targets energy proportionality with a focus on designing OS policies and mechanisms for power management. Most of this work presents hardware level optimizations that manipulate processor speed mechanisms such as DVFS ~\cite{10.5555/2523721.2523732,10.1145/381677.381702,cpufreq_governor,4273098,packandcap,10.1109/MICRO.2006.8,1598114,10.1145/1629911.1629926,4658633,4343825,10.1109/IGCC.2011.6008552,10.1145/1241601.1241609, slowdownorsleep,4228267, mootaz}, processor power limiting mechanisms such as RAPL~\cite{intel_rapl, heracles, SmoothOperator,oldi-pegasus, Dynamo,PerAppPower,powercap}, and idle power states~\cite{cpuidle_policy,peafowl, udpm,6983037,dreamweaver, pacingtoidle} (c-states) by applying feedback control mechanisms and relying on activity models. The authors of ~\cite{heracles} and ~\cite{PerAppPower} go a step further, exploring and characterizing the interference of co-located latency-critical versus best-effort tasks and high versus low CPU demand tasks when subject to energy tuning via DVFS and RAPL. In doing so, they highlight limitations in using hardware features alone for power management. Similarly, the authors of ~\cite{hotpower2008, 7349225} identify a need to step away from relying entirely on hardware solutions and focusing instead on software optimizations, such as VM migration controllers for power management of an ensemble of nodes. Previous works have advocated for full-system and hardware optimizations for energy~\cite{slowdownorsleep,powernap}, our work builds on their observations and assert that the OS itself plays a big role as well. 

The previous research efforts present significant energy savings from well designed dynamic policies and carefully chosen static configurations, however, we are driven to explore the space beyond current findings with a focus on unveiling the role of the OS in exploiting activity and idleness and also by introducing interrupt delay as an additional knob in this exploration. We find that this exploration is timely given the range of work on optimizing OS paths for performance, from NIC driver mechanisms~\cite{flexnic, affinityaccept, network-latency} to the network stack~\cite{mtcp, sandstorm, network-latency} and the dataplane~\cite{arrakis, ebbrt, shenango, zygos, ix}. Our work was also influenced by previous work in energy efficiency by slowing down both the networking and processor: $\mu$DPM~\cite{udpm} is a application-level policy for memcached to delay request processing and maximize idle periods where deep sleep states can then be utilized, in ~\cite{10.5555/2338816.2338822} the authors combined bandwidth limiting in Cray clusters and scaling processor frequency to reduce energy use of HPC applications. In contrast to $\mu$DPM, we use a hardware register on the NIC to induce batching as this can be commonly found in commercial NICs. Lastly, we are the first to conduct such an in-depth study with a baremetal specialized OS.

\section{Conclusion}
\label{sec:dis}
In this study, we conducted tens of thousands of experimental runs and accumulated over 5 TB of data. Our data also includes fine-grained timeline of various hardware/software statistics at a per-interrupt granularity (not heavily discussed in paper). This paper only scratches the surface of what can be distilled from the data and we plan to open source both the tools and data so that others can use this methodology to enable the systems research community to explore and frame performance results in context of energy as advocated by Mudge et al~\cite{917539}.

\bibliographystyle{ACM-Reference-Format}
\bibliography{references}

\appendix

\section{Mathematical Framework}
\label{sec:model}

A byproduct of our detailed analysis via this wealth of data collected is that we have begun to create a useful framework for analyzing and exploring energy and performance impacts accounting for OS behavior. We believe this framework will be used as a backbone towards more automated optimization of hardware and OS settings. In this section we briefly summarize our work in using our request time-line breakdown (Figure~\ref{fig:timeline}) to develop a mathematical framework that can be used to explain and explore software and hardware effects.  In particular, we show how we model an open loop setting, with an arrival rate of $\lambda$, and explore the impacts of changes in the instruction path length and composition.  A more detailed discussion of the framework and how it can be used can be found in Appendix~\ref{sec:appendix}.   

Assuming a setting where the service time is less than or equal to the time between two requests, we define:

$\boxed{\delta t = t_{\text{detect}} + t_{\text{osreq}} + t_{\text{app}} + t_{\text{idlepolicy}} + t_q} = \frac{1}{\lambda}$

$\delta t$ = time between the arrivals of two consecutive requests and the remaining terms directly reflect the time-line components.  

We group together the three terms that have a clear DVFS dependence and define
$t_{\text{work}}$ as $t_{\text{osreq}} + t_{\text{app}} + t_{\text{idlepolicy}}$.  Excluding detection and any quiescent time, defining $t_{\text{latency}}$ as $t_{\text{detect}} + t_{\text{work}}$ for the total service time or latency for a request.


Similarly the total energy consumed during the inter-arrival time, $\delta t$ is:

$\boxed{E = P_\text{detect} t_{\text{detect}} + P_{\text{work}} \left[t_{\text{osreq}} + t_{\text{app}} + t_{\text{idlepolicy}}\right] + P_q t_q} = P_\text{detect} t_{\text{detect}} + P_{\text{work}} t_{\text{work}} + P_q t_q$


An important aspect of our model is our physically motivated abstraction of a processor's DVFS setting, $\Delta$.  While it is a single value we model its ability to have an independent impact on time (as a possible function of frequency) and power (as a possible function of voltage and frequency).  Specifically, we posit 
$t_{\text{work}}$ and $P_{\text{work}}$ as follows:

$t_{\text{work}} = A\frac{N_i}{\Delta^{1+\alpha}}$ and $P_{\text{work}} = B \Delta^{2+\beta}$

where A, B are constants of proportionality, $N_i$ $=$ the total number of instructions and $\alpha$, $\beta$ are real-valued constants that describe the dependence on DVFS.  This allows us, through $\alpha$ and $\beta$, to explore effects in which different instruction mixes of the software, are affected in different ways with respect to time and energy by DVFS settings.  



\begin{figure}
\centering
\includegraphics[width=.8\columnwidth]{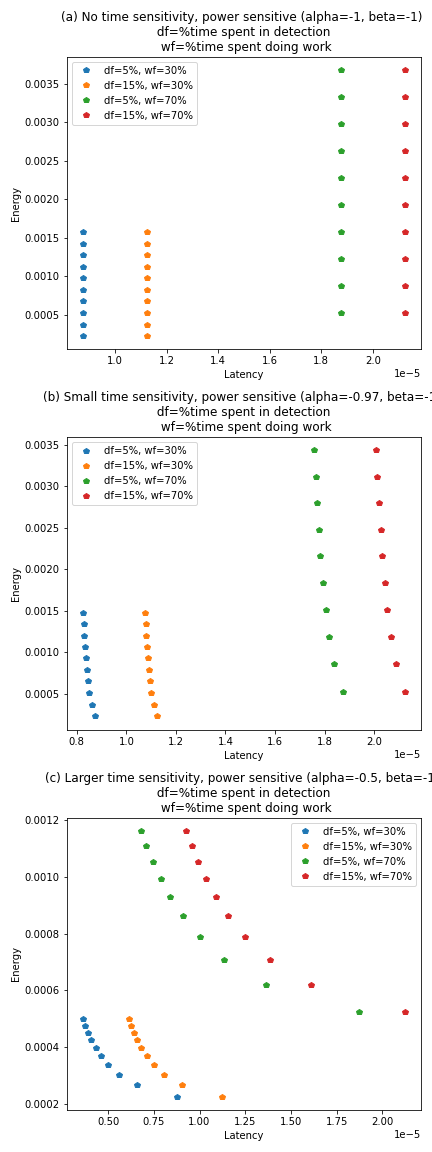}
\vspace*{-6mm}
\caption[]{Simulated Plots. Note: The X and Y axis values are not realistic values but rather simulated values.}
\label{fig:simplots}
\vspace{-0.25in}
\end{figure}

As an example of the use of the framework to generate simulated plots similar to figures shown above, in figure~\ref{fig:simplots}, plot a) shows that when instructions are less affected by frequency one expects to see a vertical structure.  Plots b) and c) show that as the paths are composed of instructions that are more frequency sensitive DVFS changes result in a more curved structure in the energy vs latency.  As expected slowing the instructions starts to affect latency in these instruction mixes.   Additionally in all three plots one sees that delaying interrupts via ITR, for a busy fraction, increases latency with not much improvement in energy.  Remembering that for this configuration deep sleep has been configured to use zero energy.  In section~\ref{sec:mcd} we will examine a scenario that arises in practice that displays similar behavior.

\newpage
\section{\bf{Appendix:}  Mathematical Framework}
\label{sec:appendix}

Under the assumption that the various time-segments in figure~\ref{fig:timeline} don't overlap, one can break down a full cycle from one request to another into disjoint time-intervals:

\begin{equation}
\delta t = t_{\text{detect}} + t_{\text{osreq}} + t_{\text{app}} + t_{\text{idlepolicy}} + t_q = \frac{1}{\lambda}
\label{eq:time}
\end{equation}

where, $\delta t$ = time between the arrivals of two consecutive requests. The other terms represent the various intervals in figure~\ref{fig:timeline}.

Each term is treated as a deterministic and fixed quantity (dependent on workload, hardware, OS, parameters) as opposed to a random variable following some underlying probability distribution. This is sufficient for a qualitative treatment but a full quantitative treatment would treat these terms as part of a probabilistic graphical model.

Note that the interval $\delta t$ is the interarrival time which is given by the reciprocal of the arrival rate (in queries/requests per unit time), $\lambda$. We will work in the regime where $\lambda$ is low enough that each request is processed before the next request arrives. While the treatment above is for the open-loop setting, this restriction also applies to the closed loop setting where, by construction, the interarrival interval always exceeds the time spent processing a request.

To map this timeline decomposition to our experimental setup, we can group together some of the terms to define:

$$t_{\text{work}} = t_{\text{osreq}} + t_{\text{app}} + t_{\text{idlepolicy}}$$

$$t_{\text{latency}} = t_{\text{work}} + t_{\text{detect}}$$

to get:

$$\delta t = t_{\text{detect}} + t_{\text{work}} + t_q = t_{\text{latency}} + t_q$$

Intuitively, $t_{\text{work}}$ is the time spent on processing the request outside the detection phase and outside any quiescent time, $t_q$, and $t_{\text{latency}}$ is the time spent both in the detection phase and on processing for a given request.

Since the total time, $\delta t$ is fixed (=$\frac{1}{\lambda})$, this implies that the quiescent time is,

$$t_q = \left[\frac{1}{\lambda} - (t_\text{work} + t_{\text{detect}})\right]^+$$

where $[x]^+ = \max(x,0)$ i.e. $[x]^+$ is the positive part of $x$. 

In other words, if the arrival rate $\lambda$ is small enough, there is an opportunity for the processor to enter a quiescent state ($t_q > 0$) but as the arrival rate increases, the time processing the request, $t_\text{work}$ exceeds the inter-arrival gap leading to requests accumulating in the queue. 

As stated above, these relationships also applies to the closed-loop case with the additional constraint that the arrival rate and thus the interarrival gap is no longer independent of $t_\text{work}$.

Given this time decomposition, one can compute the total energy consumed for each request as follows:

\begin{equation}
  E = P_\text{detect} t_{\text{detect}} + P_{\text{work}} \left[t_{\text{osref}} + t_{\text{app}} + t_{\text{idlepolicy}}\right] + P_q t_q 
\label{eq:energy}
\end{equation}
The assumption is that there are three power regimes, one each for the detection phase, the work phase and the quiescent phase, respectively.

For the open-loop case, since we are studying energy, $E$ vs latency, $t_{\text{latency}}$ plots for various itr ($t_{detect}$) and DVFS values, we need to posit the dependence of these terms on DVFS. Suppose the workload needs $N_i$ instructions. One would expect $t_\text{work}$ to scale as:

$$t_{\text{work}} \propto \frac{N_i}{f}$$

where f = CPU frequency. Of course, there might be deviations from this behavior and one can posit a power law dependence,

$$t_{\text{work}} = A\frac{N_i}{f^{1+\alpha'}}$$

where A is a constant of proportionality and $\alpha'$ is an arbitrary parameter. $\alpha'$ = 0 would fit the baseline case where time scales inversely with frequency. Since we control DVFS and not frequency directly, we can change this to

$$t_{\text{work}} = A\frac{N_i}{\Delta^{1+\alpha}}$$

where $\Delta$ = the chosen DVFS value and $\alpha$ is some scaling power that can be inferred from data. The other time values don't depend on DVFS in this simple model (although that assumption can be added in a straightforward way).

The total energy consumed depends on various power values which in turn can depend on DVFS. Here, we posit that $P_{\text{work}}$ has a power law dependence on DVFS, $\Delta$. To motivate this, the power consumed by a processor scales as:

$$P \propto V^2 f$$

where V = the operating voltage and f = CPU frequency. DVFS scales both voltage and frequency but not necessarily in a linear way. The general power law assumption is parameterized by a second parameter, $\beta$, as follows:

$$P_{\text{work}} = B \Delta^{2+\beta}$$

where B is a constant of proportionality and $\beta$ can be unrestricted and is meant to be inferred from the data. Depending on the exact setup, it is possible that $P_{\text{detect}}$ also scales with DVFS and in that case, we will set $P_{\text{detect}} = P_{\text{work}}$.

At a qualitative level, the two relationships,

\begin{equation}
    \delta t = t_{\text{detect}} + t_{\text{work}} [=\frac{AN_i}{\Delta^{1+\alpha}}] + t_q
\end{equation}

\begin{equation}
    E = P_\text{detect} t_{\text{detect}} + P_{\text{work}}[=B\Delta^{2+\beta}] t_{\text{work}}[=\frac{AN_i}{\Delta^{1+\alpha}}] + P_q t_q
\end{equation}

with the requirements that $\delta t = \frac{1}{\lambda}$ or equivalently, $t_q = \left[\frac{1}{\lambda} - [t_\text{work}+t_\text{latency}) \right]^+$ can be used to plot the behavior of energy consumed vs time (latency, total run-time) for various values of $\alpha$ and $\beta$.

We can plot some energy, $E$ vs latency, $t_{\text{latency}}$ curves numerically for the case:

$$P_{\text{detect}} = P_q = 0W$$

$$P_{\text{work}} = P_{\text{static}} + P_{text{min}} \Delta^{2+\beta} = 10W + 20W \Delta^{2+\beta}$$

For a fixed interarrival time, $\delta t$, we assign a fraction $f_{\text{detect}}$ to the detection phase and a maximum fraction $f_{\text{work}}^{\text{max}}$ to the work to get:

\begin{equation}
\begin{split}
    t_{\text{latency}} &= t_{\text{detect}} + t_{\text{work}} \\
    &= f_{\text{detect}}\delta t + \frac{f_{\text{work}}^{\text{max}}\delta t}{\Delta^{1+\alpha}}  \\
\implies & \boxed{\frac{t_{\text{latency}}}{\delta t} = f_{\text{detect}} + \frac{f_{\text{work}}^{\text{max}}}{\Delta^{1+\alpha}}}
\end{split}
\end{equation}

and

\begin{equation}
\begin{split}
    E &= P_\text{detect} t_{\text{detect}} + P_{\text{work}}  t_{\text{work}} + P_q t_q \\
    &= P_{\text{work}}  t_{\text{work}} \\
    &= (10W + 20W \Delta^{2+\beta}) (f_{\text{detect}}\delta t + \frac{f_{\text{work}}^{\text{max}}\delta t}{\Delta^{1+\alpha}}) \\
    \implies &\boxed{\frac{E}{1W \delta t} = (10 + 20 \Delta^{2+\beta})(f_{\text{detect}} + \frac{f_{\text{work}}^{\text{max}}}{\Delta^{1+\alpha}})}
\end{split}
\end{equation}

\begin{figure*}
\centering
\includegraphics[width=1.1\textwidth]{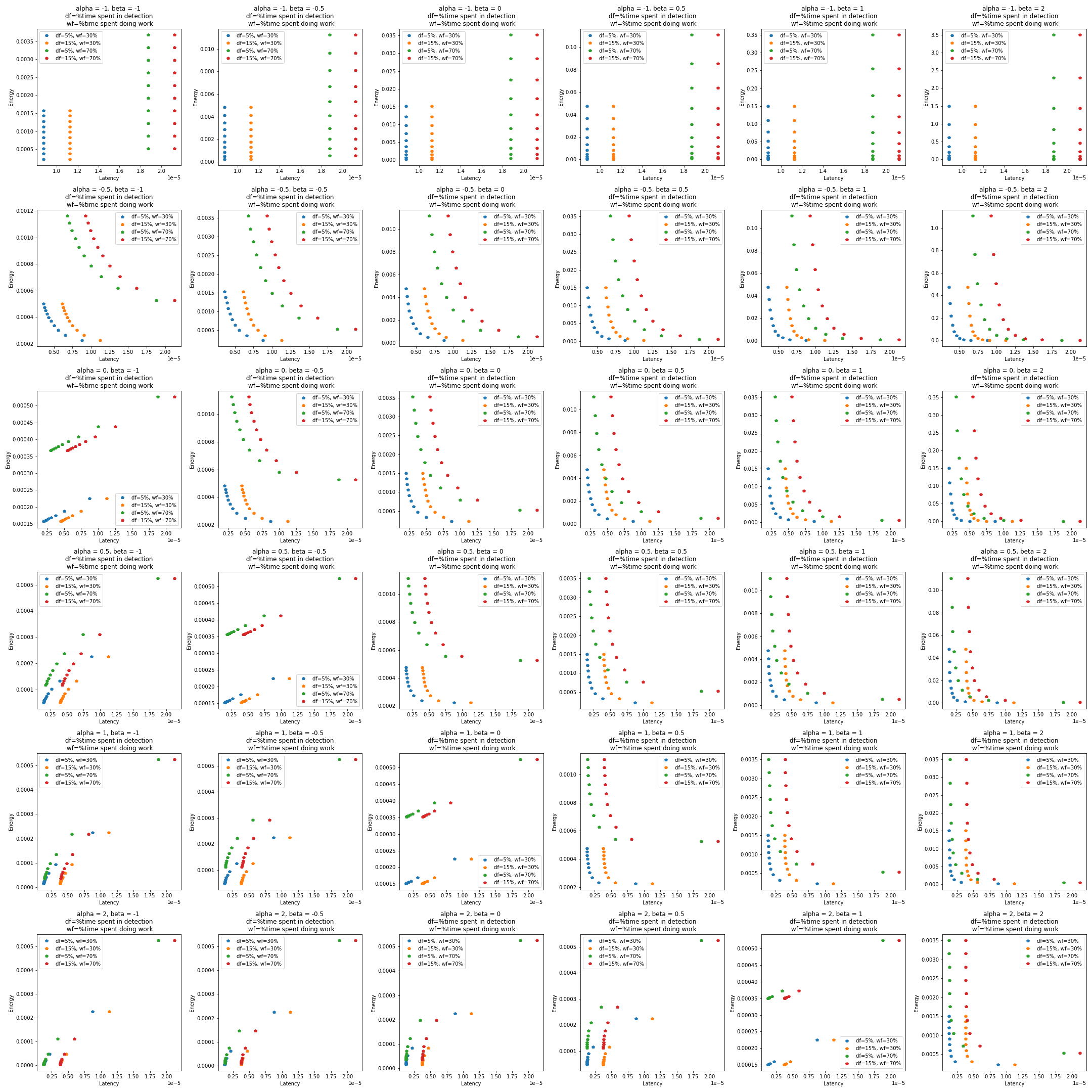}
\caption[]{Simulated Energy vs Latency Curves for Open-Loop Workloads. Power sensitivity to DVFS increase across columns and time/latency sensitivity to DVFS increases across rows}
\label{fig:simplotsfull}
\end{figure*}

For various choices of detection loads ($0 \leq f_{\text{detect}} \leq 1$), maximal work loads ($0 \leq f_{\text{work}}^{\text{max}} \leq 1$), time-scaling ($\alpha$), power-scaling ($\beta$), we can plot energy (E) vs latency ($t_{\text{latency}}$) curves as DVFS ($\Delta$) varies. An example is shown in figure \ref{fig:simplotsfull}. Power sensitivity to DVFS, $\beta$ increases across columns (left to right) and time sensitivity to DVFS, $\alpha$ increases across rows (top to bottom).

While these qualitative plots indicate the ability of the model to replicate curves observed in real data, a natural next step would be to perform fits to infer the values of the model parameters including $\alpha$ and $\beta$ and reason about them in terms of both the workload application and the underlying OS structure.

\end{document}